\documentclass[%
reprint, hidelinks,
superscriptaddress,
hypertext,
showkeys,
showpacs,
 nofootinbib,
 nobibnotes,
 amsmath,amssymb,
 aps,
prc,
floatfix,
]{revtex4-2}

\usepackage{graphicx}% Include figure files
\usepackage{dcolumn}% Align table columns on decimal point
\usepackage{bm}% bold math
\usepackage{isotope}
\usepackage{orcidlink}
\usepackage{upgreek}
\usepackage{longtable}
\usepackage[caption=false]{subfig}
\usepackage{booktabs}
\usepackage{tabularx}
\usepackage{tablefootnote}
\usepackage{threeparttable}
\usepackage{epigraph}
\usepackage{comment}
\usepackage{tikz}
\usetikzlibrary{automata,arrows,positioning,calc}
\usepackage[]{hyperref}% add hypertext capabilities
\hypersetup{
  colorlinks   = true, %Colours links instead of ugly boxes
  urlcolor     = magenta, %Colour for external hyperlinks
  linkcolor    = magenta, %Colour of internal links
  citecolor   =  cyan %Colour of citations
}
\begin{document}

\preprint{APS}

\title{Nuclear $\gamma$-Ray Cascades as Markov Processes}% Force line breaks with \\

\author{Athanasios Psaltis \orcidlink{0000-0003-2197-0797}}
\email{thanassis.psaltis@smu.ca}
\affiliation{Department of Astronomy \& Physics, Saint Mary’s University, Halifax, NS B3H 3C3, Canada}

\date{\today}% It is always \today, today,
             %  but any date may be explicitly specified
\begin{abstract}
A framework for computing $\gamma$-ray feeding probabilities in nuclear decay schemes based on absorbing Markov chains is presented.  
In this approach, excited nuclear states are treated as transient states and long-lived levels as absorbing states, allowing feeding fractions to be obtained exactly from the transition matrix. 
Experimental uncertainties are propagated via Monte Carlo sampling from Dirichlet distributions, which naturally maintains the physical constraint of unit normalization for branching-ratio vectors.
This framework is applied to the key $E_r= 92~$~keV resonance in the $\isotope[25][]{Mg}(p,\gamma)\isotope[26][]{Al}$ reaction ($E_x = 6398$~keV), which governs the production of $\isotope[26][]{Al}$ in hydrogen-burning environments. 
Combining multiple experimental datasets within a Hierarchical Bayesian framework, a ground-state feeding probability of $f_0 = 0.68 \pm 0.06~(1\sigma) \pm 0.13~(2\sigma)$ is found, and  for the first time the dominant $\gamma$-decay transitions contributing to its uncertainty are identified.
The formalism reproduces traditional cascade calculations while providing analytic sensitivity information and a transparent uncertainty decomposition. 
This approach offers a general and computationally efficient tool for propagating nuclear-structure uncertainties to astrophysical reaction rates and can be readily extended to other nuclei.
\end{abstract}

%\begin{description}
%\item[Usage]
%Secondary publications and information retrieval purposes.
%\item[Structure]
%You may use the \texttt{description} environment to structure your abstract;
%use the optional argument of the \verb+\item+ command to give the category of each item. 
%\end{description}

%\keywords{Suggested keywords}%Use showkeys class option if keyword
                              %display desired
\maketitle

%\tableofcontents

\section{Introduction}
\label{sec:1}

The de-excitation of atomic nuclei through $\gamma$-ray emission is a fundamental process in nuclear physics, underpinning both spectroscopic studies of nuclear structure and the determination of reaction rates in astrophysical environments~\cite{Eberth2008, Blumenfeld2013, Iliadis2015}.
In complex decay schemes, where an excited state can populate multiple lower-lying levels through a network of branched transitions, the resulting cascade involves a large number of interconnected paths.
The calculation of feeding probabilities and the propagation of experimental branching-ratio uncertainties in such systems is nontrivial~\cite{Gilmore2024}.
Conventional approaches often treat transitions independently or rely on linearized error propagation, which can violate normalization constraints and become increasingly inaccurate as the level density and cascade complexity increase.
A rigorous framework that treats the decay network as a complete, constrained system is therefore required.

A key example with direct astrophysical implications is the radionuclide $\isotope[26][]{Al}$ ($t_{1/2} = 7.2 \times 10^5$~a)~\cite{Prantzos1996}. 
Its characteristic 1.809 MeV $\gamma$ ray, first detected by the HEAO-3 satellite~\cite{Mahoney1984}, provides direct evidence of ongoing nucleosynthesis in the Galaxy and has since been mapped with high precision by COMPTEL and INTEGRAL/SPI~\cite{Diehl2006, Pleintinger2023}. 

The production and destruction of $\isotope[26][]{Al}$ in hydrostatic and explosive burning environments -- including massive stars, core-collapse supernovae, Wolf–Rayet stars, AGB stars, and classical novae -- remain subjects of experimental and theoretical investigation~\cite{Iliadis2011, Laird2023}.
In explosive hydrogen burning, the principal production channel is the radiative proton capture reaction $\isotope[25][]{Mg}(p,\gamma)\isotope[26][]{Al}$. 
Its thermonuclear reaction rate depends sensitively on the $\gamma$-decay branching ratios of resonant states above the proton threshold~\cite{Iliadis2026}. 
Of particular importance is the ground-state feeding probability $f_0$, which determines the fraction of $\gamma$-decays populating the long-lived ground state rather than the short-lived isomeric state at $\mathrm{E_x} = 228$~keV ($t_{1/2} = 6.35$~s). 
Accurate knowledge of $f_0$ is therefore essential for reliable $\isotope[26][]{Al}$ nucleosynthesis yield predictions and for interpreting the observed Galactic $\gamma$-ray flux.

At temperatures below $T \approx 0.4$~GK, the ground and isomeric states of $\isotope[26][]{Al}$ do not reach thermal equilibrium and must be treated as separate species in reaction network calculations~\cite{Ward1980,Runkle2001}. 
At higher temperatures, thermal coupling between the two states modifies the effective lifetime and abundance evolution. 
In either temperature regime, the fraction of $\isotope[25][]{Mg}(p,\gamma)\isotope[26][]{Al}$ reactions feeding the ground state remains a key nuclear input.

In the temperature range $T = 0.07 - 0.15$~GK, the thermonuclear reaction rate is dominated by the resonance at $\mathrm{E_r} = 92$~keV, corresponding to the $\mathrm{E_x} = 6398$~keV excited state in $\isotope[26][]{Al}$~\cite{Sallaska2013, Iliadis2026}. 
Experimental determinations of the $\gamma$-decay scheme of this excited state have been reported over several decades \cite{Champagne1986, Endt1987, Endt1988, Kankainen2021, Lotay2022, Zhang2023}. 
Although modern measurements have significantly improved the spectroscopy, notable differences among datasets persist. 
Because $\gamma$ decay proceeds through multi-step cascades, the mapping from individual branching ratios to the ground-state feeding probability $f_0$ is inherently nonlinear and involves correlations between transitions.
For this reason, discrepancies in branching ratios and their uncertainties cannot, in general, be reliably propagated through simple linear error analysis.

Despite extensive experimental effort, the impact of these differences on the inferred $f_0$ value has not been assessed within a unified statistical framework. 
Bayesian statistical methods have been successfully applied to numerous problems in nuclear astrophysics, including thermonuclear reaction rate calculations~\cite{Iliadis2016}, analyses of experimental spectroscopy and transfer reaction data~\cite{Rodgers2021,Marshall2020, Sun2023}, $R$-matrix calculations~\cite{Skowronski2025}, and statistical model parameter estimation~\cite{Chalil2024, Marshall2025}. 
In contrast, statistically rigorous uncertainty propagation through complex $\gamma$-ray cascade networks has received comparatively little attention.
In this work, the $\gamma$ decay network of $\isotope[26][]{Al}$ is modeled as an absorbing Markov Chain and the ground-state feeding probability $f_0$ as an absorbing-state probability is computed. 
This approach enables for the first time a consistent treatment of the full $\gamma$-decay scheme and allows branching-ratio uncertainties to be propagated through Monte Carlo sampling of experimental data.
In addition, the dominant $\gamma$-decay transitions that control the uncertainty in $f_0$ are identified.
The method provides a transparent and reproducible way to assess the astrophysical impact of $\gamma$-branching uncertainties.

%%%%%%%%%%%%%%%%%%%%%%%%%%%%%%%%%%%%%%%%%%%%%%%%%%%%%%%%%%%%%%%
% Paper structure
%%%%%%%%%%%%%%%%%%%%%%%%%%%%%%%%%%%%%%%%%%%%%%%%%%%%%%%%%%%%%%%
The paper is organized as follows. Section~\ref{sec:theory} introduces the formalism, Section~\ref{sec:analysis} describes its application to $\isotope[26][]{Al}$, Section~\ref{sec:results} presents the results, and Section~\ref{sec:conclusions} summarizes the conclusions and outlook.
%%%%%%%%%%%%%%%%%%%%%%%%%%%%%%%%%%%%%%%%%%%%%%%%%%%%%%%%%%%%%%%

\section{Theoretical Framework}
\label{sec:theory}

Gamma-ray cascades in atomic nuclei can be formulated as absorbing Markov chains~\cite{Norris1997}, in which excited nuclear levels are treated as transient states and long-lived levels (\textit{e.g.}, ground or isomeric states) as absorbing states.

Let the full transition probability matrix be written in canonical form:
\begin{equation}
P =\begin{bmatrix}
Q & R\\
0 & I
\end{bmatrix}
\end{equation}
where $Q$ contains transition probabilities among transient states, $R$ contains probabilities for transitions from transient to absorbing states, and $I$ represents the absorbing states.

\begin{figure}[ht!]
\begin{center}
    \includegraphics[width=.2\textwidth]{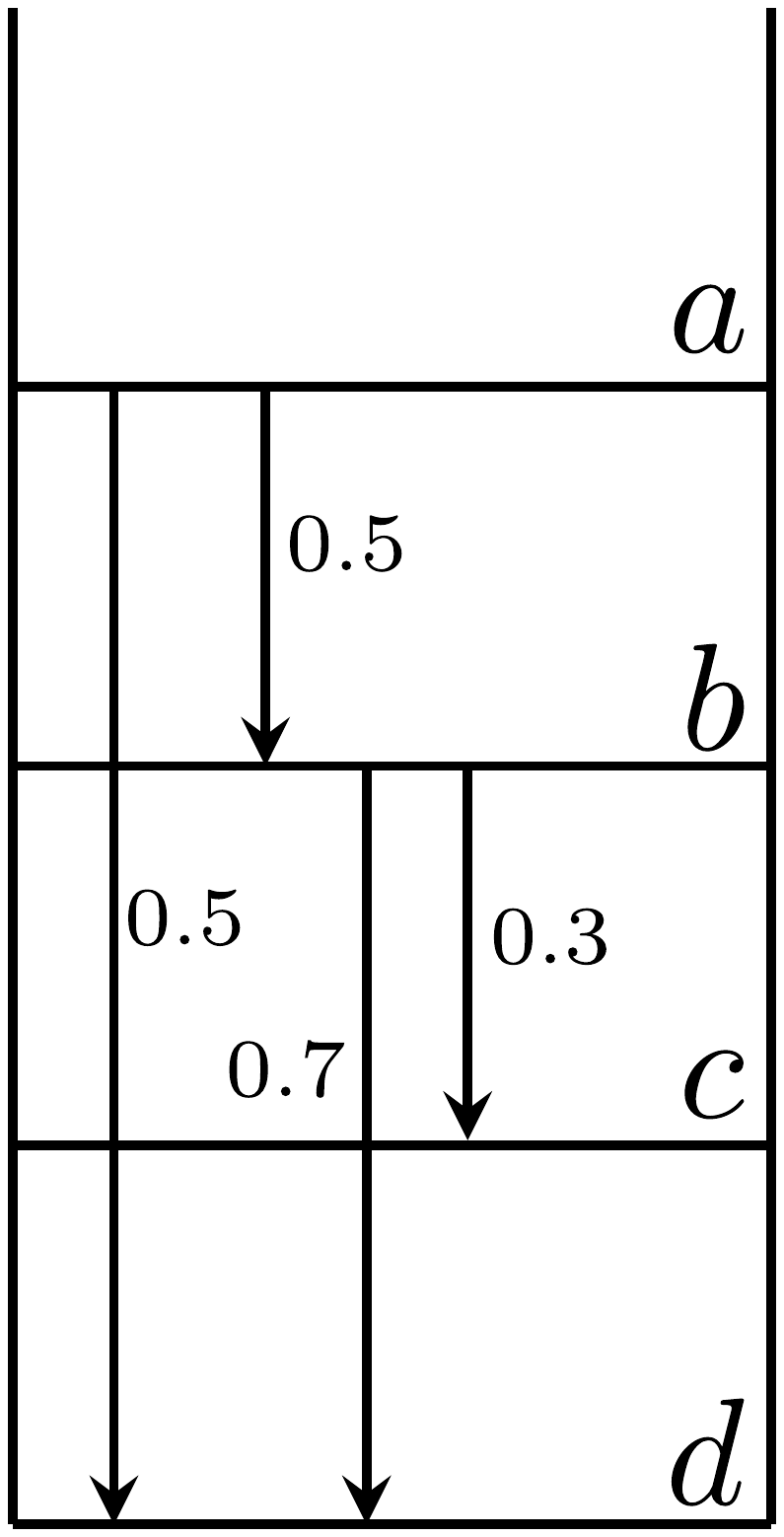}
	\begin{tikzpicture}[->, >=stealth', auto, semithick, node distance=2.5cm,]
	\tikzstyle{every state}=[fill=white,draw=black,thick,text=black,scale=0.85]
	\node[state]    (A)                     {$a$};
	\node[state]    (B)[right of=A]   {$b$};
	\node[state]    (C)[right of=B]   {$c$};
	\node[state]    (D)[right of=C]   {$d$};
	\path
    (A) edge[bend left,below]   node{$0.5$}	(D)
    (A) edge[bend left,below]   node{$0.5$}	(B)
	(B) edge[bend left,below]	node{$0.3$}	(C)
    (B) edge[bend right,below]	node{$0.7$}	(D)
	(C) edge[loop right]		node{$1$}	(C)
	(D) edge[loop right]		node{$1$}	(D);
	\end{tikzpicture}
\end{center}
\caption{(Top) A simplified decay scheme with the branching ratios for each $\gamma$-transition. (Bottom) Example of a Markov chain transition diagram for the same decay scheme. Note that states $c$ and $d$ in this example are absorbing.}
\label{fig:1}
\end{figure}

$P$ is a stochastic matrix with non-negative entries whose rows sum to unity, ensuring conservation of probability.
The transition matrix can be equivalently represented as a directed graph, as illustrated in Figure~\ref{fig:1} for a simplified $\gamma$-decay scheme of four states. 
The decay scheme can then be described by a directed acyclic graph (DAG)~\cite{Lipsky2022}, since there are no $\gamma$ ray loops, which would violate energy conservation.

For that example in Figure~\ref{fig:1}, the transition matrix is:
\begin{equation}
    P =\begin{bmatrix}
0 & 0.5 & 0 & 0.5 \\
0 & 0 & 0.3 & 0.7\\
0 & 0 & 1 & 0 \\
0 & 0 & 0 & 1
\end{bmatrix}
\end{equation}

Under an ordering of states by decreasing excitation energy, the transition matrix is typically upper triangular, reflecting the downward character of $\gamma$ decay.
The absorbing states correspond to levels that do not $\gamma$-decay further within the model space.

For absorbing Markov chains, the expected number of visits to transient states before absorption is given by the fundamental matrix~\cite{Chung1967}:

\begin{equation}
    N = (I-Q)^{-1}
\end{equation}
For a physically realistic decay scheme, all transient states ultimately connect to an absorbing state, ensuring that the spectral radius -- the maximum of the absolute values of its eigenvalues -- of $Q$ is less than unity and that $(I-Q)^{-1}$ exists.

The matrix of absorption probabilities is given by:
\begin{equation}
    B = NR
\end{equation}

Each element $B_{ij}$ gives the probability that the system, starting in transient state $i$ will eventually be absorbed in absorbing state $j$\footnote{Here, indices $i$, $k$, and $l$ denote transient states, while $j$ labels absorbing states.}.
Thus, the ground-state feeding probability $f_0$ for an initial state $i$ is:

\begin{equation}
    f_0 = B_{i,0}
\end{equation}
where $B_{i,0}$ denotes the column corresponding to the ground state. 
The absorbing-chain formalism yields feeding probabilities in a single matrix inversion, which is both exact and computationally efficient.

\begin{figure}[ht!]
   \centering
   \includegraphics[width=.5\textwidth]{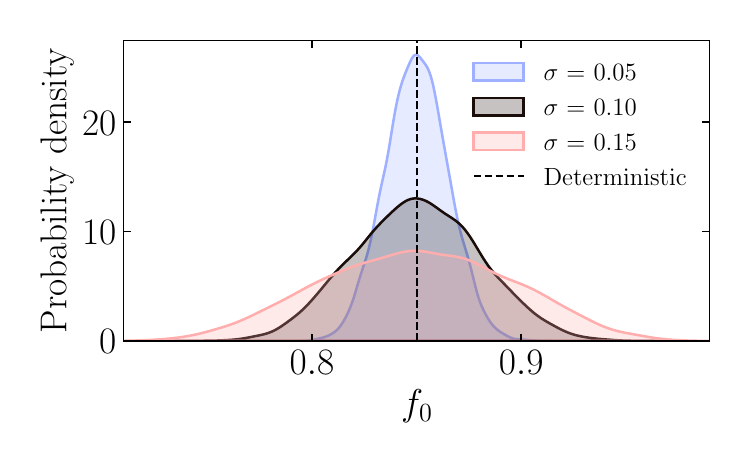}
    \caption{(Color online) Probability density function of $f_0$ for the state $a$ of the presented example.
    The vertical line indicates the deterministic calculation ($f_0=0.85$). The effect of three different uncertainties of the primary transition (5, 10, and 15\%) are shown in the resulted $f_0$ distributions.}
\label{fig:f0_example}
\end{figure}

For the example shown in Figure~\ref{fig:1}, the transient and absorbing submatrices are:
\[
Q =\begin{bmatrix}
0 & 0.5  \\
0 & 0  
\end{bmatrix} ,    R =\begin{bmatrix}
0 & 0.5  \\
0.3 & 0.7  
\end{bmatrix},
\]
where states $a$ and $b$ are transient, while $c$ and $d$ are absorbing. 
The corresponding fundamental matrix is:
\[
N = (I-Q)^{-1} = \begin{bmatrix}
1 & 0.5  \\
0 & 1  
\end{bmatrix},
\]
indicating that a cascade starting from state $a$ visits state $b$, on average, half a time before absorption.
The absorption probability matrix is then:
\[
B = NR = \begin{bmatrix}
0.15 & 0.85  \\
0.30 & 0.70  
\end{bmatrix}.
\]
The first row shows that a cascade initiated in state $a$ has a probability of 0.15 to terminate in absorbing state $c$ and 0.85 to terminate in absorbing state $d$, while the second row gives the corresponding probabilities for state $b$.
For this example, the feeding probability to state $d$ is simply:
\[
f_0 = B_{a,d}= 0.85,
\]
which is the deterministic result shown later in Figure~\ref{fig:f0_example}.

The example illustrates the physical interpretation of the absorbing-chain formalism: $Q$ describes propagation through excited states, $R$ represents direct feeding into long-lived states, $N$ gives the expected number of visits to each excited state before absorption, and $B$ provides the total feeding probabilities including all possible cascade paths. 
The feeding fraction is therefore equivalent to a first-passage absorption probability in a directed decay graph.

Uncertainties in branching ratios are encoded by sampling the rows of $P$ using Dirichlet distributions.
This choice ensures that sampled branching ratios remain normalized and naturally incorporates correlations between transitions within a given decay branch.
For each sampled matrix $P_i$, we compute:
\begin{equation}
    B_i = (I-Q_i)^{-1} R_i
\end{equation}
The distribution of $f_0$ is then obtained from the ensemble of $B_i$.

Figure~\ref{fig:f0_example} shows a calculation of the $f_0$ probability density function for state $a$ for different assumed uncertainties in the primary transitions, $a \rightarrow b$ and $a \rightarrow d$ (5, 10, and 15\%). 
The deterministic result of $f_0 = 0.85$ is reproduced, and the uncertainty is properly propagated through a Monte Carlo method by sampling the uncertainty using a Dirichlet distribution. 
The Dirichlet distribution preserves normalization constraints and ensures non-negative transition probabilities, which would not be possible if Gaussian uncertainties were chosen.
The concentration parameters $\boldsymbol{\alpha}$ are chosen such that the mean of each transition probability reproduces the reported branching ratio and the variance matches the quoted experimental uncertainty, ensuring consistency between the sampling procedure and published data.
The selection of $\boldsymbol{\alpha}$ is discussed in more detail in Section~\ref{sec:analysis}.

Although the full uncertainty propagation is performed via Monte Carlo sampling, analytic insight can be obtained through first-order uncertainty propagation. 
To leading order, the variance of $f_0$, assuming \textit{independent branching ratio uncertainties}, can be written as:
\begin{equation}\label{eqn:7}
    \mathrm{Var}(f_0) \approx \sum_{i,j} \left(\frac{\partial f_0}{\partial p_{ij}}\right)^2 \sigma_{ij}^2
\end{equation}
where $\sigma_{ij}$ is the uncertainty associated with transition $p_{ij}$.
In the absence of reported covariance matrices, branching ratio uncertainties are currently treated as independent.

Using matrix identities for absorbing Markov chains, the derivative with respect to a transient–transient transition $Q_{kl}$ is:
\begin{equation}\label{eqn:8}
    \frac{\partial f_0}{\partial Q_{kl}} = N_{ik} B_{l0},
\end{equation}
while for a transient state:
\begin{equation}\label{eqn:9}
    \frac{\partial f_0}{\partial R_{k0}} = N_{ik}
\end{equation}

These expressions have a clear physical interpretation; 
the factor $N_{ik}$ represents the expected number of visits to state $k$ starting from the initial state $i$, while $B_{l0}$ is the probability that a cascade starting from state $l$ ultimately will feed the ground state. 
Thus, a transition contributes strongly to the variance of $f_0$ if the cascade frequently passes through its parent level, and the daughter level has a strong connection to the ground state.

For each transition, we define the partial variance contribution as:
\begin{equation}
\Delta \mathrm{Var}_{kl}= \left(\frac{\partial f_0}{\partial P_{kl}}\right)^2 \sigma_{kl}^2    
\end{equation}
The fractional contribution to the total variance is then:
\begin{equation}\label{eqn:11}
    F_{kl} = \frac{\Delta \mathrm{Var}_{kl}}{\sum_{m,n} \Delta \mathrm{Var}_{mn}}
\end{equation}
This quantity provides a quantitative ranking of the contributions of individual $\gamma$-decay transitions to the total uncertainty in $f_0$.
Importantly, this decomposition is independent of Monte Carlo sampling and follows directly from the analytic structure of the absorbing chain.

The variance decomposition reveals that not all large branching ratios are equally influential. 
A weak transition can dominate the uncertainty if it lies on a frequently traversed cascade path with a strong connection to the ground state. 
On the other hand, even sizable branching ratios may contribute negligibly if the cascade rarely passes through the corresponding level.
A transition from state $k$ to $l$ contributes strongly to 
the $f_0$ uncertainty if the cascade frequently visits $k$ (large $N_{ik}$) and if decays from $l$ have high probability of eventually reaching the ground state (large $B_{l0}$).

This analysis identifies the specific $\gamma$-decay transitions whose improved measurement would most effectively reduce the uncertainty in $f_0$, providing a quantitative prioritization for future experimental studies.

Although current experimental data lack reported covariances between branching ratios, the Markov chain framework is readily extensible to incorporate such correlations. 
If future measurements provide covariance information, the Monte Carlo sampling can be adapted by replacing the independent Dirichlet draws with multivariate normal sampling in a transformed space (\textit{e.g.}, logistic-normal). 
The analytic sensitivity coefficients Equations~\ref{eqn:8},\ref{eqn:9} would then combine with the full covariance matrix via Equation~\ref{eqn:7} to partition variance contributions including correlation effects. 
This extensibility ensures the framework remains relevant as experimental reporting standards evolve in the future.

\begin{table}[ht]
\centering
\begin{threeparttable}
\caption{Primary $\gamma$-branching ratios for the
$E_r=92$~keV resonance of
$\isotope[25]{Mg}(p,\gamma)\isotope[26]{Al}$
($E_x=6398$~keV). Values are given as decimal fractions; $\cdots$ denotes a non-observation of that transition.}
\label{tab:BRs}
\begin{tabular}{ccccc}
\toprule
$E_x$ (keV) & Set 1 & Set 2 & Set 3 & Set 4 \\
\midrule
5142 & 0.09(3) & 0.0069(12) & 0.0112(39) & 0.07(2) \\
3403 &0.04(2) & $\cdots$ & $\cdots$ & $\cdots$ \\
3160 & 0.76(8) & 0.355(17) & 0.543(11) & 0.53(2) \\
2070 & $<0.02$ & 0.512(23) & 0.154(4) & 0.31(2) \\
1850 & 0.05(2) & $\cdots$ & $\cdots$ & $\cdots$ \\
1759 & $\cdots$ & 0.127(7) & 0.291(12) & 0.07(2) \\
0    & 0.04(2) & $\cdots$ & $\cdots$ & 0.02(1) \\
\bottomrule
\end{tabular}

\begin{tablenotes}
\footnotesize
\item Set 1: \citet{Champagne1986}
\item Set 2: \citet{Kankainen2021}
\item Set 3: \citet{Lotay2022}
\item Set 4: \citet{Zhang2023}
\end{tablenotes}
\end{threeparttable}
\end{table}

\section{Analysis Method}
\label{sec:analysis}

\begin{figure*}[t!]
    \centering
    
    \includegraphics[width=0.85\textwidth]{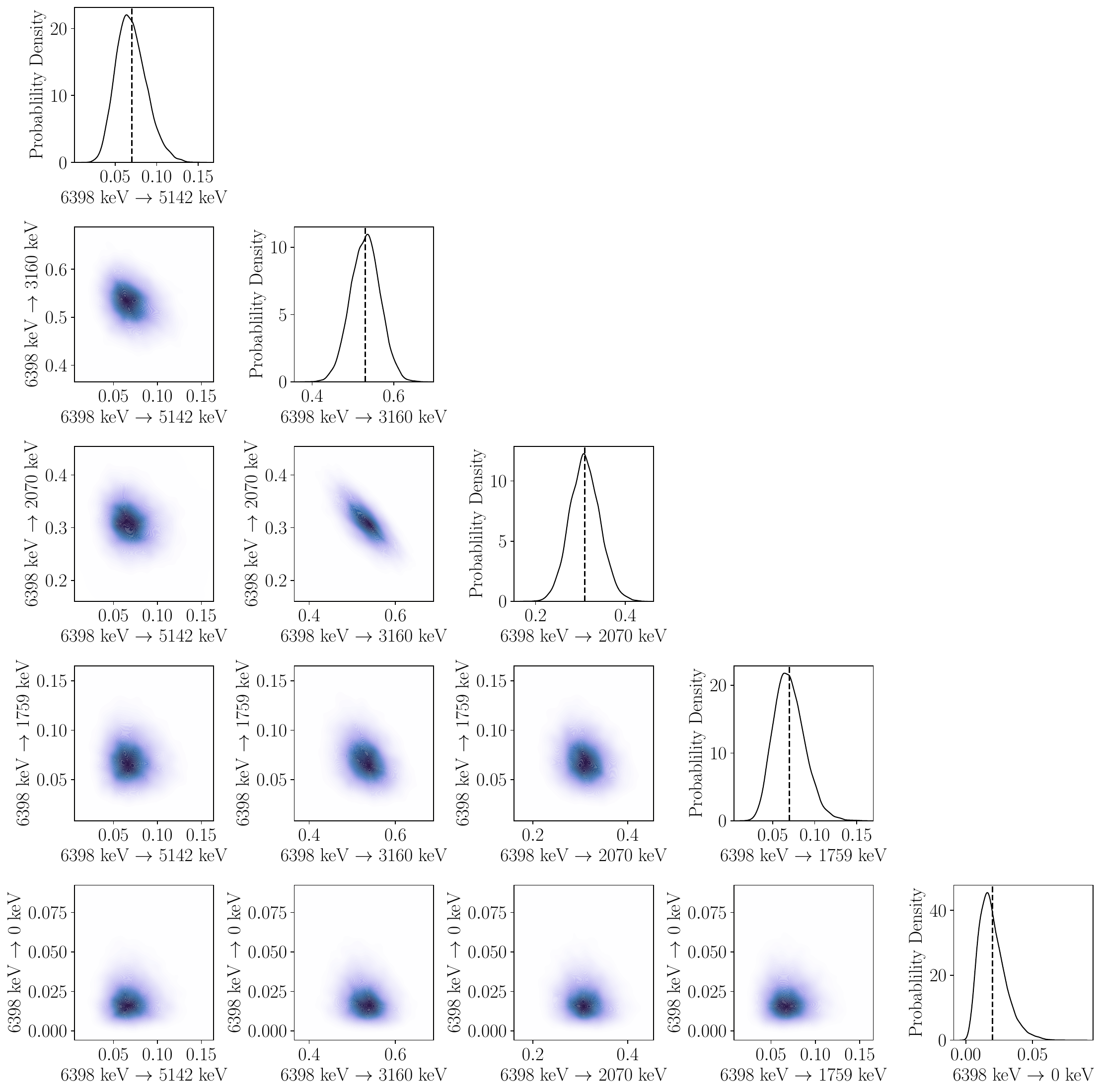}
    \caption{(Color online) Joint Probability Density Functions (PDFs) for the $\gamma$-transitions of the $E_x$= 6398~keV state using the branching ratios from~\citet{Zhang2023}. 
    The black vertical dashed lines correspond to the reported branching ratios. 
    The anticorrelation of the two strongest transitions is evident, suggesting a proper sampling.}
    \label{fig:sampling}
\end{figure*}

In this section, the absorbing Markov-chain framework is applied to compute the ground-state feeding probability $f_0$ for excited states in $\isotope[26][]Al$, using literature values of $\gamma$-decay branching ratios and their associated uncertainties. 
The goal is to provide a consistent and statistically robust determination of $f_0$, with particular focus on the astrophysically important $E_x=6398$~keV state.

77 excited states ($E_x = 228 - 6816$~keV) in $\isotope[26][]{Al}$ and their $\gamma$-ray branching ratios from the $A= 26$ data evaluations~\citet{Endt1990,Basunia2016} are used to reproduce the results of~\citet{Endt1987} which provide the only published $f_0$ values for excited states in $\isotope[26][]{Al}$. 
Specifically for the astrophysically interesting state at $E_x = 6398$~keV ($E_r= 92~$~keV resonance in $\isotope[25][]{Mg}(p,\gamma)\isotope[26][]{Al}$), calculations of the $f_0$ are performed using different literature values for the $\gamma$-ray branching ratios reported in Refs.~\cite{Champagne1986, Kankainen2021, Lotay2022, Zhang2023} (see Table~\ref{tab:BRs}) and this is discussed in more detail in Section~\ref{sec:results}.

Compiling all the relevant data from the literature yields a ``master'' transition matrix with all the different $\gamma$-transitions.
Care must be taken in the identification of closely spaced levels. A notable example is the ``2069 keV'' state reported in Refs.~\cite{Kankainen2021, Zhang2023}, which corresponds to the $E_x=2069.47$~keV level and should not be confused with the $E_x=2068.86$~keV level. 
In earlier evaluations~\cite{Endt1990}, these are labeled as ``2070'' and ``2069'' keV, respectively. 
This distinction is important, as the two states have significantly different feeding properties ($f_0 \approx  0.22$ vs. $f_0=1$).
Throughout this work, ``2070 keV'' refers to the level at $E_x=2069.47$~keV following the convention of ~\citet{Endt1990}.

An uncertainty matrix $U$ of the same dimensions is constructed to store the reported experimental uncertainties $\sigma_i$ of the $\gamma$-branching ratios.
The uncertainty matrix is used to properly sample the transition matrix $P$.
Figure~\ref{fig:sampling} shows an example of the sampling process for the $E_x =$ 6398~keV state using the branching ratio data from~\citet{Zhang2023}.
Each branching ratio is sampled according to a Dirichlet distribution, which ensures that probability is conserved.

Each row of the transition matrix, corresponding to the branching ratios from a given excited state, is modeled as a Dirichlet-distributed random vector. 
The Dirichlet sampling is applied independently to each row of the transition matrix, corresponding to the assumption that experimental uncertainties are \textit{uncorrelated} between different initial states.
The concentration parameters are defined as $\boldsymbol{\alpha}=\kappa \mathbf{p}$, where $\mathbf{p}$ is the vector of reported branching ratios and $\kappa$ controls the overall precision.
The value of $\kappa$ is determined by matching the Dirichlet marginal variances, $\text{Var}(p_i)=p_i(1-p_i)/(\kappa+1)$, to the reported experimental uncertainties $\sigma_i^2$. 
In practice, $\kappa$ is estimated for each channel and the median positive solution is adopted to ensure numerical stability.
This approach preserves normalization of the branching ratios and consistently incorporates the induced correlations between transitions. 
Without this treatment, independent sampling of branching ratios leads to biased estimates of $f_0$.

The convergence of the Monte Carlo sampling was verified by examining the stability of the variance estimate of $f_0$. 
For $N=5,000$ samples, the expected relative statistical uncertainty of the variance estimator is $\approx \sqrt{2/(N-1)} \approx 2\%$, consistent with the observed fluctuations. 
Increasing the number of samples to $N=15,000$ does not significantly change the inferred uncertainty, indicating convergence.

For each of the 5,000 samples, $B_i$ were computed and then an ensemble of $f_0$ was obtained which was used to extract the median value and the $2\sigma$ uncertainty (approximately 2\textsuperscript{nd} and 98\textsuperscript{th} percentiles).
In addition, the variance decomposition formalism (Equation~\ref{eqn:11}) was applied to identify the individual $\gamma$-decay transitions that dominate the uncertainty in $f_0$, providing direct guidance for future experimental efforts.

All calculations were performed on a standard desktop workstation. 
A Monte Carlo calculation with 5,000 samples for a single state (\textit{e.g.} the $E_x$= 6398~keV) required approximately 8~s using a single CPU core.

\section{Results}
\label{sec:results}

\begin{figure}[ht!]
    \centering
    \includegraphics[width=.46\textwidth]{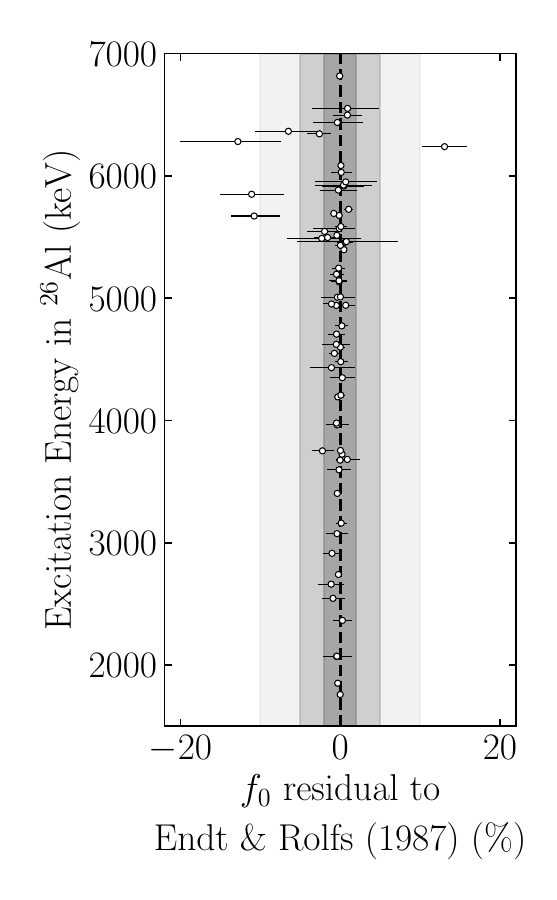}
    \caption{Comparison of ground-state feeding probability $f_0$ for excited states in $^{26}$Al. 
    Residuals are shown relative to the evaluation of~\citet{Endt1987}.
    The three grey shaded horizontal bands represent the $\pm 2\%$, $\pm 5\%$, and $\pm10\%$ residual ranges, respectively. 
    All states shown agree within 10\% at the $2\sigma$ level. 
    See Figure~\ref{fig:6398} for the $\mathrm{E_x}$= 6398~keV state.}
    \label{fig:f0_Endt}
\end{figure}

Figure~\ref{fig:f0_Endt} compares the ground-state feeding probabilities $f_0$ obtained with the present framework to the evaluation of~\citet{Endt1987}, using the nuclear data compilation of~\citet{Basunia2016}.
For the majority of states, the Markov-chain calculation reproduces the literature values to within $\approx 2\%$ (relative difference), while all states agree within $10\%$ when accounting for the propagated $2\sigma$ uncertainties.

At excitation energies above $\approx 5.5$~MeV, the deviations increase systematically. 
This behavior reflects the increasing level density and the corresponding growth in the number of accessible decay pathways, leading to more complex multi-step cascades.
In this regime, small differences in individual branching ratios can propagate nonlinearly through the cascade and produce amplified shifts in $f_0$.
The Markov-chain formalism provides a robust framework for treating these complex decay networks, avoiding the accumulation of errors inherent in sequential multiplicative propagation methods.

Several states reported in~\citet{Endt1987} have incomplete $\gamma$-decay schemes (\textit{e.g.}, 5598, 6086, 6198, 6270, 6399, and 6414~keV). 
For these cases $f_0$ were not computed, as the missing transitions prevent a complete and self-consistent construction of the transition matrix.

\begin{figure}[ht!]
    \centering
\includegraphics[width=.5\textwidth]{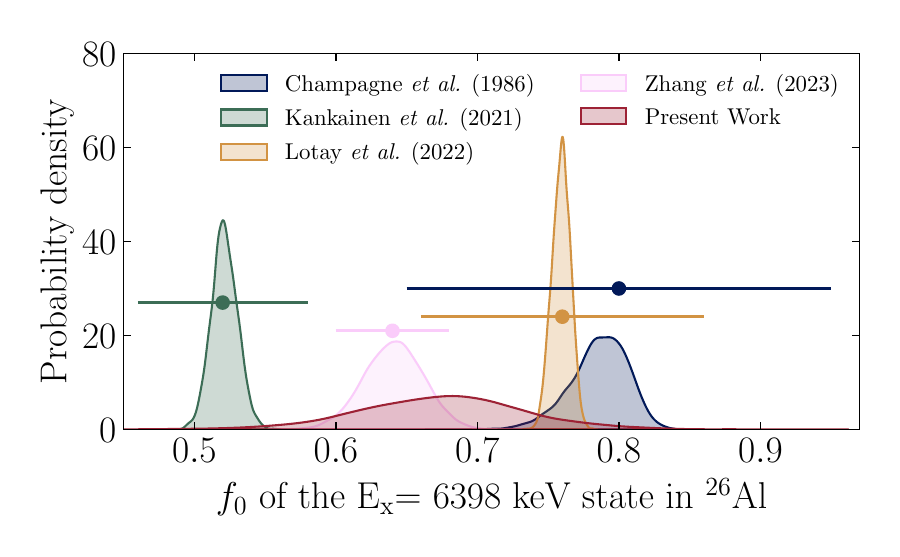}
    \caption{(Color online) Probability density functions (PDFs) of the ground-state feeding probability $f_0$ for the $\mathrm{E_x}$= 6398~keV in $\isotope[26][]{Al}$, derived from different experimental $\gamma$-ray transition datasets (\citet{Champagne1986, Kankainen2021, Lotay2022, Zhang2023}) and propagated through the Markov-chain framework. The points with error-bars correspond to the reported $f_0$ values in the respective publications.
    The maroon curve shows the PDF produced using the Markov-Chain PDFs and a Hierarchical Bayesian Model.
    }
    \label{fig:6398}
\end{figure}

Figure~\ref{fig:6398} shows the probability density functions (PDFs) of $f_0$ for the $E_x=6398$~keV state derived from the different experimental datasets. 
Table~\ref{tab:BRs} illustrates the substantial disagreement among the published branching-ratio measurements, with several primary transitions differing by factors of two or more despite modern high-resolution spectroscopy.
The distributions are obtained by propagating the reported branching ratios and uncertainties through the Markov-chain framework.
While the original publications report central values and uncertainties, the full statistical propagation of branching-ratio uncertainties to $f_0$ is not treated explicitly. 
The present framework provides this propagation self-consistently, yielding well-defined PDFs for each dataset.
Most datasets produce approximately symmetric distributions, although the result of~\citet{Champagne1986} exhibits a pronounced low-$f_0$ tail. 
In general, the uncertainties obtained here are smaller than those quoted in the literature, reflecting the consistent treatment of correlations and normalization constraints. 
The result of~\citet{Zhang2023} is an exception, with a comparable uncertainty of $\approx 4\%$.

To derive a recommended value for $f_0$, the results from the four experimental datasets~\cite{Champagne1986, Kankainen2021, Lotay2022, Zhang2023} are combined, accounting for the evident tension between them.
A direct multiplication of the individual probability densities would lead to an artificially over-constrained estimate, because the reported distributions are not fully consistent.
Instead, a hierarchical Bayesian approach is adopted in which an additional systematic uncertainty term $\sigma_{\rm sys}$ represents the unknown inter-experimental scatter.

The posterior distribution is defined as:
\begin{equation}
P(f_0|\{D_i\}) \propto
\int
\left[
\prod_{i=1}^{4}
P(D_i|f_0,\sigma_{\rm sys})
\right]
\pi(\sigma_{\rm sys})
\,d\sigma_{\rm sys},
\end{equation}
For each experimental dataset, the Monte Carlo samples are first converted into a kernel density estimate (KDE). 
For a given value of $\sigma_{\rm sys}$, each KDE is convolved with a Gaussian kernel of width $\sigma_{\rm sys}$, thereby broadening the likelihood to account for possible systematic differences between experiments. 
Rather than selecting a single value of $\sigma_{\rm sys}$, the joint posterior is evaluated over a grid of $\sigma_{\rm sys}$ values using a uniform prior -- $\sigma_{\rm sys} \sim \mathcal{U}(0,0.15)$ -- and numerically marginalize over this nuisance parameter.
The upper bound ($\sigma_{\rm sys} = 0.15$) was chosen to exceed the observed disagreement between the experimental datasets, and increasing it further does not modify the marginalized posterior.
The resulting posterior therefore naturally incorporates both the statistical uncertainties within each dataset and the additional uncertainty arising from their mutual disagreement.

The resulting combined value is:
\[f_0 = 0.68 \pm 0.06~(1\sigma) \pm 0.13~(2\sigma)\]
where the median and the 68\% and 95\% credible intervals (approximately 16\textsuperscript{th}–84\textsuperscript{th} and 2\textsuperscript{th}–98\textsuperscript{th} percentiles, respectively) are reported.

This result is consistent with the values of~\citet{Lotay2022},~\citet{Champagne1986}, and~\citet{Zhang2023} within $1\sigma$, and with~\citet{Kankainen2021} within $2\sigma$. 
The lower $f_0$ value reported by~\citet{Kankainen2021} can be traced to an enhanced population of the 2069~keV state, as also noted by~\citet{Lotay2022}.

The value of $f_0$ directly affects the fraction of $\isotope[26][]{Al}$ produced in its ground state and therefore influences predictions of the observable 1.809 MeV $\gamma$-ray flux in the Galaxy.

\begin{figure}[ht!]
    \centering
    \includegraphics[width=.5\textwidth]{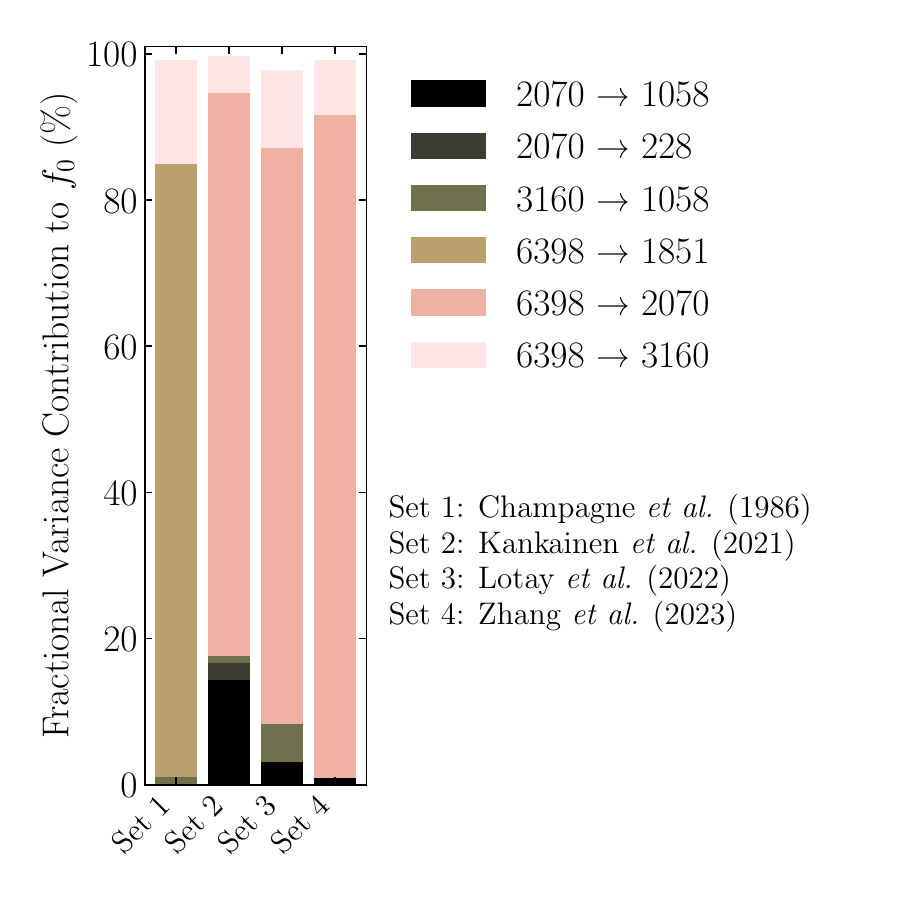}
    
    \caption{(Color online) Relative contributions of individual decay branches from the 6398~keV state to the total variance in the ground-state feeding probability $f_0$, accounting for normalization-induced correlations via a Dirichlet model.
    Only transitions with more than $\approx 1\%$ contribution to the variance have been included.}
    \label{fig:f0_var}
\end{figure}

\begin{table*}[t!]
\centering
\setlength{\tabcolsep}{2pt} % Shrink space between columns
\begin{threeparttable}
\caption{Relative contributions of individual decay branches from the 6398~keV state to the total variance of $f_0$.}
\label{tab:2}
\begin{tabularx}{\textwidth}{lccccc} % Set total width to text width
\toprule
Transition & \citet{Champagne1986} & \citet{Kankainen2021} & \citet{Lotay2022} & \citet{Zhang2023} & Note \\
\midrule
6398 $\rightarrow$ 2070 & $\ldots$ & 77.02\% & 78.81\% & 90.7\% & \small Top transition in 3 datasets \\
6398 $\rightarrow$ 3160 & 14.23\% & 5\% & 10.62\% & 7.48\% & \small Top 2 in 3 datasets; top 1 in one \\
2070 $\rightarrow$ 1058 & $\ldots$ & 14.38\% & 3.10\% & 0.96\% & \small Top 3 in 3 datasets \\
6398 $\rightarrow$ 1851 & 83.90\% & $\ldots$ & $\ldots$ & $\ldots$ & \small Top transition in 1 dataset \\
\bottomrule
\end{tabularx}
\end{threeparttable}
\end{table*}

In addition to determining $f_0$, the present framework allows a decomposition of its variance into contributions from individual $\gamma$-decay transitions.
Figure~\ref{fig:f0_var} shows the fractional contributions to the total variance for the four datasets, including only transitions contributing more than $\approx 1\%$, and are compiled in Table~\ref{tab:2}.
The results are largely consistent across datasets, with the $6398 \rightarrow 2070$ transition dominating the uncertainty in three out of four cases.
This analysis provides a direct and quantitative identification of the transitions that most strongly influence $f_0$, offering a clear prioritization for future experimental studies.

\section{Discussion and Conclusions}
\label{sec:conclusions}

This work presents an absorbing Markov-chain framework for the calculation of $\gamma$-ray feeding probabilities that provides a fully consistent treatment of branching-ratio uncertainties and their correlations.

Applied to $\isotope[26][]{Al}$, the method reproduces previous cascade calculations while enabling a statistically rigorous propagation of experimental uncertainties.
For the astrophysically important $E_x=6398$~keV state, a recommended ground-state feeding probability of
\[f_0 = 0.68 \pm 0.06~(1\sigma) \pm 0.13~(2\sigma)\]
was derived from a hierarchical combination of the available experimental datasets.
The remaining spread between measurements reflects significant inter-experimental inconsistencies, which constitute the dominant source of uncertainty in $f_0$.
A key result of this work is the decomposition of the total variance of $f_0$ into contributions from individual $\gamma$-decay transitions.
This analysis identifies a small number of dominant branches, most notably the $6398 \rightarrow 2070$ transition, as the primary drivers of the uncertainty.
This provides, for the first time, a quantitative and experimentally actionable prioritization of which transitions should be targeted to reduce the uncertainty in $f_0$.

The present framework is general and can be readily applied to other nuclei with complex decay schemes.
Its computational efficiency and transparent statistical structure make it well suited for incorporation into future nuclear data evaluations.
Extensions to include thermally populated states will allow direct coupling to stellar reaction rates, enabling the systematic propagation of nuclear-structure uncertainties into nucleosynthesis calculations and $\gamma$-ray observables.

\section*{Acknowledgements}
I acknowledge the support of the Natural Sciences and Engineering Research Council of Canada (NSERC) under grant SAPIN-2026-00045. 
I thank Christian Iliadis and Richard Longland for useful discussions, and the referee for suggestions that improved the manuscript.

 \section{Data Availability}
All analysis scripts and input data required to reproduce the figures and results of this work are available in a public repository released under the AFL-3.0 License (Zenodo doi: \href{https://doi.org/10.5281/zenodo.21359901}{10.5281/zenodo.21359901})~\cite{zenodo}.

\bibliographystyle{apsrev4-2}
\bibliography{bibliography}% Produces the bibliography via BibTeX.

\end{document}